\documentclass[draftcls, 12pt,onecolumn]{IEEEtran}
\usepackage{float}
\usepackage{cite}

\usepackage[cmex10]{amsmath}
\usepackage{amssymb}
\usepackage{amsthm}
\usepackage{mathrsfs}
\usepackage{bm}
\usepackage[mathscr]{eucal}
\usepackage{amssymb,amsmath,amsthm,amsfonts,latexsym}
\usepackage{amsmath,graphicx,bm,xcolor,url}
\usepackage{graphicx}
\usepackage{latexsym}
\usepackage{CJK}
\usepackage{indentfirst}
\usepackage{geometry}
\usepackage{psfrag}
\usepackage{setspace}
\usepackage{algorithmic}
\usepackage{algorithmic, cite}
\usepackage{algorithm}
\usepackage{array}
\usepackage{mdwmath}
\usepackage{mdwtab}
\usepackage{eqparbox}
\usepackage{url}
\usepackage{epstopdf}
\usepackage{epsfig,epsf,psfrag}
\usepackage{fixltx2e}
\usepackage{verbatim}
\usepackage{textcomp}
\usepackage{psfrag} 

\usepackage{caption}
\usepackage{subcaption}

\theoremstyle{plain}
\newtheorem{mythe}{Theorem}
\theoremstyle{remark}
\newtheorem{mylem}{Lemma}
\theoremstyle{plain}
\newtheorem{mydef}{Definition}
\theoremstyle{remark}
\newtheorem{mypro}{Proposition}
\theoremstyle{plain}

\theoremstyle{remark}
\newtheorem{Remark}{Remark}
\theoremstyle{remark}

\theoremstyle{remark}

\theoremstyle{remark}

\theoremstyle{remark}

\theoremstyle{remark}

\catcode`~=11 \def\UrlSpecials{\do\~{\kern -.15em\lower .7ex\hbox{~}\kern .04em}} \catcode`~=13

\allowdisplaybreaks[4]

\newcommand{\calC}{\mathcal{C}}

\newcommand{\calN}{\mathcal{N}}

\newcommand{\calR}{\mathcal{R}}

\newcommand{\calX}{\mathcal{X}}
\newcommand{\calY}{\mathcal{Y}}

\newcommand{\ba}{\mathbf{a}}
\newcommand{\bA}{\mathbf{A}}

\newcommand{\be}{\mathbf{e}}

\newcommand{\bh}{\mathbf{h}}

\newcommand{\bI}{\mathbf{I}}

\newcommand{\bw}{\mathbf{w}}

\newcommand{\bX}{\mathbf{X}}

\newcommand{\bbC}{\mathbb{C}}

\newcommand{\bbE}{\mathbb{E}}

\DeclareMathAlphabet{\mathbsf}{OT1}{cmss}{bx}{n}
\DeclareMathAlphabet{\mathssf}{OT1}{cmss}{m}{sl}

\DeclareSymbolFont{bsfletters}{OT1}{cmss}{bx}{n}
\DeclareSymbolFont{ssfletters}{OT1}{cmss}{m}{n}
\DeclareMathSymbol{\bsfGamma}{0}{bsfletters}{'000}
\DeclareMathSymbol{\ssfGamma}{0}{ssfletters}{'000}
\DeclareMathSymbol{\bsfDelta}{0}{bsfletters}{'001}
\DeclareMathSymbol{\ssfDelta}{0}{ssfletters}{'001}
\DeclareMathSymbol{\bsfTheta}{0}{bsfletters}{'002}
\DeclareMathSymbol{\ssfTheta}{0}{ssfletters}{'002}
\DeclareMathSymbol{\bsfLambda}{0}{bsfletters}{'003}
\DeclareMathSymbol{\ssfLambda}{0}{ssfletters}{'003}
\DeclareMathSymbol{\bsfXi}{0}{bsfletters}{'004}
\DeclareMathSymbol{\ssfXi}{0}{ssfletters}{'004}
\DeclareMathSymbol{\bsfPi}{0}{bsfletters}{'005}
\DeclareMathSymbol{\ssfPi}{0}{ssfletters}{'005}
\DeclareMathSymbol{\bsfSigma}{0}{bsfletters}{'006}
\DeclareMathSymbol{\ssfSigma}{0}{ssfletters}{'006}
\DeclareMathSymbol{\bsfUpsilon}{0}{bsfletters}{'007}
\DeclareMathSymbol{\ssfUpsilon}{0}{ssfletters}{'007}
\DeclareMathSymbol{\bsfPhi}{0}{bsfletters}{'010}
\DeclareMathSymbol{\ssfPhi}{0}{ssfletters}{'010}
\DeclareMathSymbol{\bsfPsi}{0}{bsfletters}{'011}
\DeclareMathSymbol{\ssfPsi}{0}{ssfletters}{'011}
\DeclareMathSymbol{\bsfOmega}{0}{bsfletters}{'012}
\DeclareMathSymbol{\ssfOmega}{0}{ssfletters}{'012}

\newcommand{\hata}{\widehat{a}}

\newcommand{\hatba}{\widehat{\ba}}
\newcommand{\hatbA}{\widehat{\bA}}

\newcommand{\hatE}{\widehat{E}}

\newcommand{\tilE}{\widetilde{E}}

\newcommand{\hath}{\widehat{h}}

\newcommand{\hatbh}{\widehat{\bh}}

\newcommand{\tiln}{\widetilde{n}}

\newcommand{\hatq}{\widehat{q}}

\newcommand{\barE}{\bar{E}}

\def\norm#1{\left\| #1 \right\|}
\def\norm2#1{\left\| #1 \right\|_2}
\def\norm22#1{\left\| #1 \right\|_2^2}

\newcommand{\eqa}{\stackrel{(a)}{=}}
\newcommand{\eqb}{\stackrel{(b)}{=}}
\newcommand{\eqc}{\stackrel{(c)}{=}}
\newcommand{\eqd}{\stackrel{(d)}{=}}

\newcommand{\lea}{\stackrel{(a)}{\le}}
\newcommand{\leb}{\stackrel{(b)}{\le}}

\newcommand{\qednew}{\nobreak \ifvmode \relax \else
      \ifdim\lastskip<1.5em \hskip-\lastskip
      \hskip1.5em plus0em minus0.5em \fi \nobreak
      \vrule height0.75em width0.5em depth0.25em\fi}
      
\title{Multi-Antenna Wireless Energy Transfer for Backscatter Communication Systems}

\author{Gang~Yang, Chin~Keong~Ho, and Yong~Liang~Guan  
\thanks{G.~Yang and Y.~L.~Guan are with the School of Electrical and Electronic Engineering, Nanyang Technological University, Singapore (e-mail:\{yang0305, eylguan\}@ntu.edu.sg).} 
\thanks{C. K. Ho is with the Institute for Infocomm Research, A$^\star$STAR, Singapore (e-mail: hock@i2r.a-star.edu.sg). }
}

\begin{document}
\maketitle 
\vspace{-1.8cm}
\begin{abstract}
We study RF-enabled wireless energy transfer (WET) via energy beamforming, from a multi-antenna energy transmitter (ET) to multiple energy receivers (ERs) in a backscatter communication system, such as RFID, where each ER (or RFID tag) reflects back a portion of the incident signal to the ET (or RFID reader). For such a system, the acquisition of the forward-channel (i.e., ET-to-ER) state information (F-CSI) at the ET is challenging, since the ERs are typically too energy-and-hardware-constrained to estimate or feed back the F-CSI. The ET leverages its observed backscatter signals to estimate the backscatter-channel (i.e., ET-to-ER-to-ET) state information (BS-CSI) directly. We first analyze the harvested energy obtained by using the estimated BS-CSI. Furthermore, we optimize the channel-training energy and the energy allocation weights for different energy beams, for weighted-sum-energy (WSE) maximization and proportional-fair-energy (PFE) maximization. For WET to single ER, we obtain the optimal channel-training energy in a semi-closed form. For WET to multiple ERs, the optimal WET scheme for WSE maximization is shown to use only one energy beam. For PFE maximization, we show it is a biconvex problem, and propose a block-coordinate-descent based algorithm to find the close-to-optimal solution. Numerical results show that with the optimized solutions, the harvested energy suffers slight reduction of less than $10\%$, compared to that obtained by using the perfect F-CSI. Hence, energy beamforming by using the estimated BS-CSI is promising, as the complexity and energy requirement is shifted from the ERs to the ET.

\end{abstract}


\section{Introduction}\label{introduction}
Recently, backscatter radio has been utilized widely, due to its low energy requirement and low monetary cost. Backscatter radio performs communication by means of reflection of incident signals rather than direct radiation,  The most prominent commercial use of backscatter radio is in radio frequency identification (RFID) applications, for identifying people or products in supply chains. For backscatter communication, the input impedance of the tag's antenna is intentionally mismatched to scatter back a portion of the incident signal. The phase and amplitude of the backscattered signal is then determined by the input impedance. By varying the antenna impedance, an RFID tag encodes digital symbols into the backscattered signal, which is then received and decoded by the RFID reader~\cite{BoyerSumit14}. The tag operates without any on-tag energy source and relies entirely on backscatter, which leads to its energy efficiency and cost effectiveness. The existing literature on communication theoretic aspect of backscatter focuses on the tag-to-reader channels, such as on multiple access techniques~\cite{ZhuYum10}, performance of space-time code~\cite{DurginTAN08}, as well as the diversity-multiplexing tradeoff for multi-input-multi-output (MIMO)~\cite{BoyerSumit14}.

Backscatter communication is also valuable to other systems built with low-power and low-cost principles, such as wireless sensor networks. The ongoing integration of various sensors on RFID tags, such as described in~\cite{Vannucci08}~\cite{AKanBaykal10}, confirms the potential for new sensor networks that use modified RFID components to transfer sensor data to the fusion center. An example of an RFID-based sensor platform is the wireless integrated sensing platform (WISP) \cite{SampleSmith07}.

However, the coverage range of RFID sensors is significantly limited by the forward channels (i.e., reader-to-tag), as only a small amount of RF energy can be harvested at the conventional RFID sensors. In the literature, the range of commercial RFID tags is improved from aspects including rectifier circuit design (see~\cite{Carvalho14} and references therein) and special waveform design~\cite{BoaventuraCarvalho13}. The waveform is designed such that the received signal frequently exceeds the threshold voltage required to turn on the rectifier circuit.

Multi-antenna techniques have been shown to be efficient for enhancing the efficiency of wireless energy transfer (WET) for traditional radio communication systems. The electromagnetic (EM) energy needs to be concentrated into a narrow beam to achieve efficient transmission of energy, referred to as {\textit{energy beamforming}}~\cite{MIMOWIPTZhang13}, as EM waves decay quickly over distances. The channel state information (CSI) is prerequisite for energy beamforming. The ERs perform channel estimation (CE) by receiving pilots sent from the ET, and then feed back the estimated forward\footnote{For consistence, the terminology of ``downlink channel'' and ``uplink channel'' in traditional radio communication systems is renamed to be ``forward channel'' and ``backward channel'', respectively. } channel (i.e., ET-to-ER) state information (F-CSI) to the ET; or the ET receives pilots sent from the ERs, and then obtains the estimated F-CSI directly by exploiting the channel reciprocity. The effect of CE and feedback on energy beamforming was studied in~\cite{YangHoGuan13, XuZhangTSP14,ZengZhangTCOM14,YangHoGuanMMIMO14}. In particular, \cite{YangHoGuan13} investigated the dynamic allocation of time resource for CE and energy resource for WET. \cite{XuZhangTSP14} studied energy beamforming by using one-bit feedback from each ER, to facilitate hardware implementation. Based on the one-bit feedback, the ET adjusts transmit beamforming and concurrently obtains improved estimates of the forward channels to all ERs. \cite{ZengZhangTCOM14} maximized the net harvested energy after subtracting the energy used for the ER sending pilots, for a point-to-point MIMO WET system. Furthermore, with the estimated F-CSI, \cite{YangHoGuanMMIMO14} optimized the throughput for a massive MIMO system powered by WET.


{For backscatter communication systems, energy beamforming can also be used to achieve efficient WET. Multiple antennas are deployed at the ET (or RFID reader) to perform energy beamforming toward ERs (or RFID tags). However, unlike traditional radio communication systems, the acquisition of the F-CSI at the ET is challenging, since the ERs are typically too energy-constrained to perform CE or feedback, and also may not have specific hardware built for CE nor feedback.}

Instead, the ET may in fact leverage its observed backscatter signal to {estimate the backscatter-channel (i.e., ET-to-ER-to-ET) state information (BS-CSI)} solely by itself. This shifts complexity and energy requirements from the ERs to the ET. The necessary synchronization condition for CE can also be achieved more easily for the ET than for the ERs. {A proof of concept was presented in~\cite{ArnitzReynolds13}, which illustrated that the WET can be optimized by using only the power levels received at the receive antennas of the ET. However, in~\cite{ArnitzReynolds13}, the ET transfers energy to only one ER in a time period, even though other ERs may desire energy and can also potentially harvest energy concurrently. Moreover, that work did not consider the effect of CE nor the resource allocation for WET to multiple ERs concurrently.}

In this paper, based on a backscatter communication system, we consider the WET from an ET with multiple antennas to multiple ERs each with a single antenna. We assume frame-based transmissions, where each frame consists of a CE phase and a WET phase. With unknown backward channels (i.e., ER-to-ET), we first analyze the energy that can be harvested via energy beamforming by using the estimated BS-CSI. Furthermore, we optimize the resource allocation, by investigating two utility maximization problems, namely, the weighted-sum-energy (WSE) maximization and the proportional-fair-energy (PFE) maximization. The optimization variables are the channel-training energy used for CE, and the energy allocation weights for beamforming toward multiple ERs. For comparison, we consider two benchmarks, one is the ideal case of energy beamforming by using the perfect F-CSI, in which the most energy can be transferred via energy beamforming in any wireless communication system; the other is energy beamforming by using the estimated F-CSI fed back from the ERs, in a traditional radio communication system.

The main contributions of this paper are as follows:
\begin{itemize}
\item We propose a novel energy beamforming scheme by using the estimated BS-CSI, to perform WET to multiple ERs concurrently. This scheme shifts the complexity and energy requirements from the ERs to the ET, and is thus especially attractive for transferring energy to (ultra-)low-power and low-cost wireless devices that {can neither estimate channels nor send pilots or feedback}.

\item We obtain an analytical expression for the harvested energy obtained by using the estimated BS-CSI, in which the ambiguity of unknown backward channels is taken into account. We also obtain bounds on the harvested energy, which are numerically shown to be tight.

\item We obtain the optimal resource allocation schemes for both WSE maximization and PFE maximization. For the single-ER case, we obtain the optimal channel-training energy in a semi-closed form. For the multiple-ER case, to achieve WSE maximization, the optimal WET scheme is shown to use only one energy beam. For PFE maximization, we show it is a biconvex problem, and propose a block-coordinate-descent (BCD) based algorithm to find the close-to-optimal solution. Numerical results show that fairness is improved by PFE maximization.

\item {We conduct simulation studies, which show that the maximally harvested energy suffers \emph{slight} reduction of less than $10\%$ when compared to the harvested energy obtained by using the perfect F-CSI, and of about $3\%$ when compared to the net harvested energy obtained by using the estimated F-CSI in a traditional radio communication system. For the latter benchmark, the net harvested energy is the harvested energy after subtracting the energy used for sending backward pilots and feeding back the estimated F-CSI. This observation is encouraging, as almost all complexity of the ER is shifted to the ET.}
\end{itemize}

The rest of this paper is organized as follows: Section~\ref{SystemModel} presents the system model.  Section \ref{WPT_Model} analyzes the harvested energy. Section \ref{sec:Formulation} considers the resource allocation, by formulating a general utility-maximization problem. Section \ref{sec:solutions} obtains the optimal resource allocation for WSE maximization and PFE maximization. Section~\ref{Sec:Extension} gives some discussion. Section~\ref{Simulation} provides extensive numerical results. Finally, Section~\ref{Conclusion} concludes this paper.

\section{System Model} \label{SystemModel}
\begin{figure} [htbp]
\centering
\includegraphics[width=.7\columnwidth] {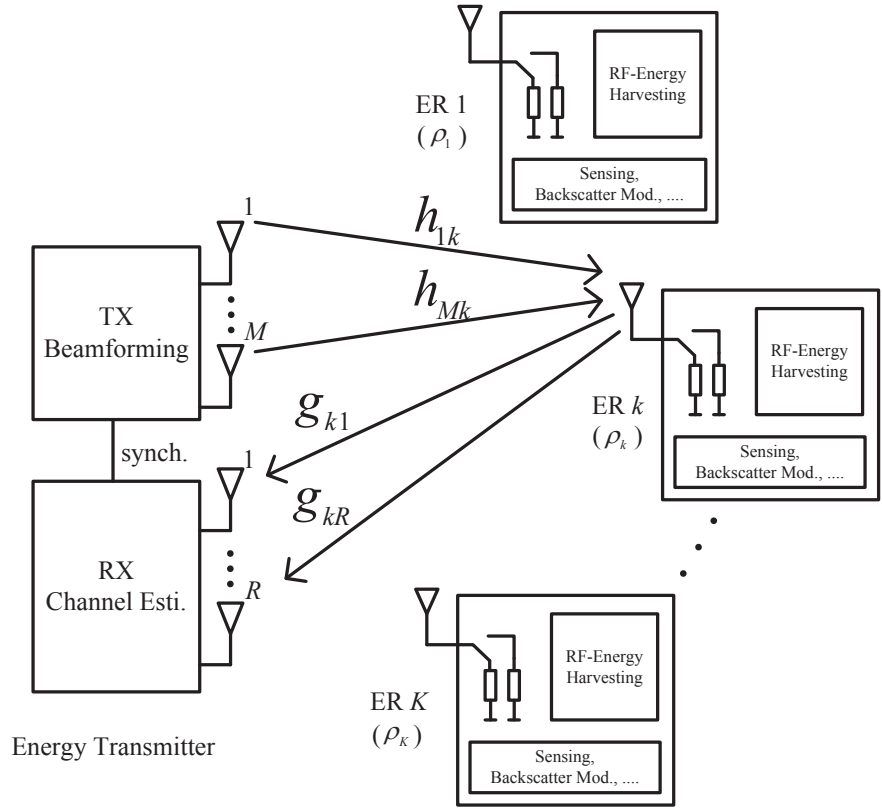}
\caption{System model}
\label{fig:Fig1}
\end{figure}

As illustrated in Fig.~\ref{fig:Fig1}, based on a backscatter communication system, we consider WET from an ET (or RFID reader) that can concurrently transmit with $M$ antennas and receive with $R$ antennas, to $K$ ERs (or RFID tags) each with single antenna. The ET can perform CE, energy beamforming and other signal processing operations. Each ER contains an RF-energy harvesting module which supplies energy for operations such as sensing, quantization and backscatter modulation. Besides, it has a switched load impedance that is connected to its antenna. By varying the antenna impedance, each ER encodes digital symbols into the backscattered signal, which is then received and decoded by the ER. Compared to traditional radio communication systems, backscatter communication is energy efficient, as it just reflects the incident signal, without generating radio signals actively. Hence, backscatter communication is alluring for (ultra-)low-power devices~\cite{BoyerSumit14}.


\begin{figure} [htbp]
\centering
\includegraphics[width=.7\columnwidth] {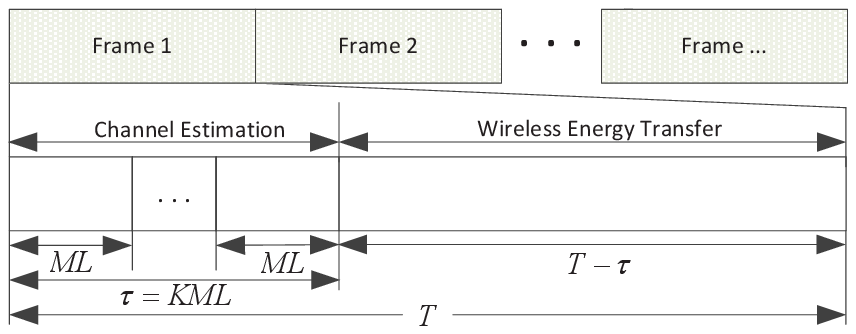}
\caption{Frame structure}
\label{fig:Fig2}
\end{figure}

We study WET via frame-based transmissions on a single frequency band. As illustrated in Fig.~\ref{fig:Fig2}, the time duration of each frame is fixed as $T$ symbol periods, which consists of the CE phase followed by the WET phase. The CE phase of $\tau$ symbol periods is further equally divided into $K$ slots, each of which consists of $ML$ successive symbol periods. That is, the CE time is $\tau=K M L$. During the CE phase, the ET sends pilot signal with power $p_1$. With the coordination\footnote{{The ET can coordinate all ERs to switch on in any arbitrary sequence. For instance, the ET sends an initialization symbol to signal the start of the CE phase, and then transmits a unique ID at the beginning of each CE slot. The corresponding ER then responds accordingly.}} of the ET, in the $k$-th CE slot, only the load impedance of the $k$-th ER is switched on to facilitate backscatter communications, with the impedance of all other ERs switched off. The pilot signal is thus backscattered by only the $k$-th ER. After receiving the backscatter signal, the ET estimates the backscatter-channel associated to the $k$-th ER. During the WET phase of $(T-\tau)$ symbol periods, the ET performs energy beamforming, and all ERs switch off the load impedance and harvest wireless energy.
Typically, the time duration of the CE phase is much shorter than the WET phase. Hence, we assume that the ERs do not harvest energy during the CE phase, for simplicity.



\subsection{Backscatter Channel}\label{sec:Channel}
The backscatter channel is modeled as a concatenation of three components, namely, forward channel (i.e., reader-to-tag), backscatter reflection coefficient, and backward channel (i.e., tag-to-reader). Let $h_{mk}$ denote the forward channel between the $m$-th transmit antenna and the $k$-th ER, and $g_{kr}$ denote the backward channel between the $k$-th ER and the $r$-th receive antenna. Denote the (long-term) path loss of the channel between the ET and $k$-th ER by $\beta_k$, which is assumed to be constant over frames and taken to be known a priori at the ET. We assume the forward and backward channels are flat Rayleigh-fading and independent, i.e., the channel coefficients $h_{mk} \sim \calC \calN (0, \beta_k)$ and $g_{kr} \sim \calC \calN (0, \beta_k)$. In the $k$-th CE slot, the load impedance of ER $k$ is switched on, and ER $k$ reflects a portion of the incident signal to the ET, which is modeled by the complex reflection coefficient $\rho_k \in \bbC$. For other ER $j$, the impedance is switched off, i.e., $\rho_j = 0,\; \forall j \neq k$. In this paper, we assume the ERs are linear backscatter devices, i.e., the reflection coefficient $\rho_k$ is fixed and does not vary with the incident power at the ERs. Without loss of generality, we assume $\rho_k=1$.

The backscatter channel associated with the $m$-th transmit antenna, the $k$-th ER and the $r$-th receive antenna is given by $a_{mkr} = h_{mk} g_{kr}$. Hence, the backscatter channel experiences double fading due to $h_{mk}$ and $g_{kr}$. For convenience of expression, we let the number of receive antenna $R=1$ and omit the subscript $r$ in the notation. The analysis can be easily extended to the case of multiple receive antennas, but is beyond the scope of this paper.

\subsection{Channel Estimation via Backscatter Signal}
The total energy for sending pilots during the whole CE phase is fixed as $\tau p_1$. We assume the ET spend equal pilot energy (i.e., $MLp_1$) for estimating the backscatter channel associated with each ER\footnote{In general, the training energy for different ERs can be adjusted depending, for example, on the long-term path loss. However, that is beyond the scope of this paper.}. Under the assumption of independent channels, from~\cite{HassibiHochwaldIT03}, the least-square (LS) estimation performance can be optimized by using the pilot matrix $\bX = [\bX_1 \; \bX_2 \ \cdots \ \bX_L]^T$ in each slot, where $\bX_l$ is an orthogonal matrix such as the Hadamard matrix or the identity matrix, with power $p_1$, i.e., $\bX_l \bX_l^H = p_1 \bI_M$. The $(\cdot)^T$ and $(\cdot)^H$ denote transpose and conjugate transpose, respectively. For convenience, we choose $\bX_l = \sqrt{p_1} \bI_M, \; l=1, 2, \cdots, L$.

When the ET sends pilots, the pilot signal can also be received directly by its own receive antenna, without going through the backscatter channel. For simplicity, we assume this undesired signal can be estimated accurately and subtracted from the received signal, as this direct channel is static and can be estimated a priori. Hence, the received signal with respect to the pilot sent from the $m$-th transmit antenna is given by
\begin{align}
  y_{mk}^{\sf CE} = L \sqrt{p_1} a_{mk} + \sum \limits_{l=1}^L n_{mkl}, \label{eq:RXsigPreamble}
\end{align}
where the ET noise $n_{mkl}$'s are independent and distributed as $\calC \calN (0, \sigma^2)$. The LS estimate for the backscatter channel with respect to the $m$-th transmit antenna and the $k$-th ER is obtained as
\begin{align}
  \hata_{mk} = \frac{y_{mk}^{\sf CE}}{L \sqrt{p_1}} = a_{mk} + \tiln_{mk}^{\sf CE}, \label{eq:Est_a}
\end{align}
where the equivalent noise $\tiln_{mk}^{\sf CE} \sim \calC \calN (0, \frac{\sigma^2}{L p_1})$. For energy beamforming, the estimate of the F-CSI $h_{mk}$ is desired. With the estimated BS-CSI $\hata_{mk}$, there is however still a remaining ambiguity due to the unknown backward channel $g_k$. We will see later the effect of the ambiguity $g_k$ on the harvested energy, in Section~\ref{sec:WET_BF_SU} and Section \ref{sec:comparison_energy}.

However, the performance of CE depends strongly on $g_k$. We define the intermediate random variable $\hath_{mk} \triangleq \hata_{mk} g_k^{-1}$, and the error $e_{mk} \triangleq \hath_{mk}- h_{mk}$. Conditioned on $g_k$, it is then standard to show that the forward channel $h_{mk}$ is conditionally distributed as
  \begin{align}\label{eq:conditional_distribution}
  h_{mk} | \hata_{mk}, g_k \sim \calC \calN \left( \frac{\beta_k \hath_{mk}}{\beta_k + \sigma_{e, k }^2 (g_k)}, \frac{\beta_k \sigma_{e, k}^2 (g_k)}{\beta_k + \sigma_{e, k}^2 (g_k)} \right),
  \end{align}
and the error is conditionally distributed as $e_{mk} \sim \calC \calN (0, \sigma_{e, k}^2 (g_k) )$, with error variance
\begin{align}
  \sigma_{e,k}^2 (g_k) = \frac{K M \sigma^2}{|g_k|^2 \tau p_1}. \label{eq:CondErrorVar}
\end{align}

\begin{Remark}[Effect of unknown backward channel on forward-channel estimation]\label{rem:effect_CE}
For the special case of $g_k = 1$ and hence $a_{mk}=h_{mk}$, the estimated F-CSI $\hath_{mk}$ is given in~\eqref{eq:Est_a}, which is exactly the same as that for a traditional radio communication system. However, for a backscatter communication system, from~\eqref{eq:CondErrorVar}, the estimation error depends on the (unknown) backward channel $g_k$, as the signal-to-noise (SNR) for CE is affected by $g_k$. Thus we have to account for the unknown $g_k$ to derive the statistics of the expected harvested energy later in  Section~\ref{WPT_Model}.
\end{Remark}

\section{Wireless Energy Transfer via Energy Beamforming}\label{WPT_Model}
In this section, we study the WET by using the estimated BS-CSI. In order to analyze how it differs from WET by using the estimated F-CSI in traditional radio communication systems, we first consider the single-ER case in Section \ref{sec:WET_BF_SU}, and analyze the effect of unknown backward channel on the harvested energy in Section \ref{sec:comparison_energy}. The case of multiple ER is studied in Section \ref{sec:WET_BF_MU}, in which we obtain bounds on the harvested energy, so as to simplify analysis in next sections.

\subsection{WET to Single ER}\label{sec:WET_BF_SU}
Instead of considering the specific case of WET to one ER, for full generality, we consider the case of WET to ER $k$. For beamforming toward only ER $k$, the $M \times 1$ transmitted signal is given by $\sqrt{p_2} \bw ({\hatba_k})$, where $p_2$ is the transmit power for WET, and $\bw(\hatba_k)$ is the beamformer depending on the estimated BS-CSI $\hatba_k$. We note that traditionally the estimated F-CSI $\hatbh_k$ is used to obtain the beamformer. The received signal by ER $k$ is written as
\begin{align}\label{eq:signal-received}
  z_k = \sqrt{p_2} \bw^T (\hatba_k) \bh_k + n_k.
\end{align}
where the noise at ER $k$ is distributed as $\calC \calN (0, \sigma_{k, 0}^2 )$. Due to the law of energy conservation with efficiency $\eta$, the RF-band energy harvested by ER $k$ during the WET phase, denoted by $E_{k}$, is assumed to be proportional to that of the received baseband signal, i.e.,
\begin{align}\label{eq:energy_per_symbol_1}
  E_{k} = E_{k} (\tau, p_1, p_2) = \eta p_2 (T-\tau) \bbE_{\bh_k, \hatba_k} \left[\left|\bw^T (\hatba_k) \bh_k \right|^2\right].
\end{align}
We assume in~\eqref{eq:energy_per_symbol_1} that the energy due to the noise at ER $k$ cannot be harvested. For convenience, we also assume $\eta=1$ in this paper. Let $(\cdot)^{\ast}$ denote the complex conjugate. The expected harvested energy in~\eqref{eq:energy_per_symbol_1} is rewritten as
  \begin{align}
    E_{k} (\tau, p_1, p_2) &= p_2 (T \!- \!\tau) \bbE_{g_k} \Big[ \bbE_{\hatba_k | g_k} \big[ \bw^T (\hatba_k) \bbE_{\bh_k | \hatba_k, g_k} \left[ \bh_k \bh_k^H\right] \bw^{\ast} (\hatba_k) \big]\Big] \nonumber \\
    &\eqa p_2 (T - \tau) \bbE_{g_k} \Big[ \bbE_{\hatba_k | g_k} \big[ \bw^T (\hatba_k) \left( \frac{\beta_k \sigma_{e, k}^2 (g_k) \bI_M}{\beta_k + \sigma_{e, k}^2(g_k)}  + \frac{\beta_k^2 \hatbh_k \hatbh_k^H}{\left(\beta_k + \sigma_{e, k}^2 (g_k) \right)^2}\right) \bw^{\ast} (\hatba_k) \big]\Big], \label{eq:energy_per_symbol_2}
  \end{align}
where the conditional correlation matrix in (a) is obtained from the conditional distribution in \eqref{eq:conditional_distribution}. Unlike traditional radio communication systems,  the harvested energy in~\eqref{eq:energy_per_symbol_2} is obtained by performing expectation over $g_k$, As noted in Remark~\ref{rem:effect_CE}, this takes into account the ambiguity of $g_k$.

%
Net, we obtain the optimal beamformer (with unknown $g_k$) in the following lemma.
\begin{mylem}\label{lemma:optimalbeamformerSU}
  The optimal energy beamformer toward a single ER $k$ is given by
  \begin{align}\label{eq:optimalbeamformerSU}
    \bw (\hatba_k) = \frac{\hatba_k^{\ast}}{\| \hatba_k\|}.
  \end{align}
The corresponding maximally harvested energy by ER $k$ is given by
  \begin{align}\label{eq:energy_per_symbol_SU_2}
    E_{k} (\tau, p_1, p_2) &= p_2 M \beta_k (T-\tau) \left( 1- \frac{M-1}{M} \bbE_{g_k}  \left[\frac{1}{\frac{\beta_k \tau p_1 |g_k|^2}{K M \sigma^2} + 1} \right] \right).
    \end{align}
  \end{mylem}

\begin{IEEEproof}
  See Appendix~\ref{appendix_proof_optimal_beamformer}.
\end{IEEEproof}

\begin{Remark}[Discussion for the special case of large number of transmit antennas $M$ or users $K$] As the product $MK$ tends to infinity, the harvested energy in~\eqref{eq:energy_per_symbol_SU_2} approaches the quantity $p_2 \beta_k (T-\tau)$. This is equivalent to the case when the wireless energy is harvested from the omnidirectional signal transmitted by the ET. This can be explained intuitively as follows. For large $M$ or $K$, a finite amount of energy has to be shared for training over all channels between the transmit antennas and the antennas of all ERs, which leads to inaccurate estimates of the BS-CSI. Hence, the beamforming gain that can be achieved for WET is very limited.
\end{Remark}

\subsection{Effect of Unknown Backward Channel on Harvested Energy} \label{sec:comparison_energy}
In this section, we analyze the effect of unknown backward channel $g_k$ on the harvested energy, still assuming WET to single ER for exposure. For radiative communication systems, the optimal beamformer is the normalized estimated F-CSI~\cite{YangHoGuan13}. From Lemma~\ref{lemma:optimalbeamformerSU}, for backscatter communication systems in which the backward channel $g_k$ is unknown, the optimal beamformer that achieves the maximally harvested energy in~\eqref{eq:energy_per_symbol_SU_2} is the normalized estimated BS-CSI. In the proof for Lemma~\ref{lemma:optimalbeamformerSU}, we show that when $g_k$ is given, the optimal beamformer is just the normalized estimated F-CSI. Intuitively, this is because $g_k$ is common for estimating all the forward channels between the ET and the ER. Similar observation was also obtained in \cite{ArnitzReynolds13}.


However, this ambiguity of backward channel $g_k$ results in a reduction of harvested energy, as shown the following proposition.
\begin{mypro}\label{pro:comparison}
Assuming the same receive SNR for CE, the harvested energy $E_{k} (\tau, p_1, p_2) $ for a backscatter communication system is upper bounded by that for a traditional radio communication system, given by
\begin{align}\label{eq:convex_upperbound}
\barE_{k} (\tau, p_1, p_2) \triangleq p_2 M \beta_k (T-\tau) \left[ 1- \frac{M-1}{M} \left(\frac{1}{\frac{\beta_k^2 \tau p_1}{K M \sigma^2} + 1} \right) \right].
\end{align}
\end{mypro}

\begin{IEEEproof}
Define the random variable $Y_k \triangleq |g_k|^2$, which follows exponential distribution. We further define the function $f(y_k)=\frac{1}{c_k y_k +1}$, where $c_k = \beta_k \tau p_1 / (K M \sigma^2)$. It can be easily checked that $f(y_k)$ is a strictly convex function of $y_k$ for $y_k>0$, hence, $\bbE_{Y_k} [f(Y_k)] \geq \frac{1}{c_k \bbE_{Y_k} [Y_k] + 1}$ with $\bbE_{Y_k} [Y_k]=\beta_k$, due to Jensen's inequality. Thus, the harvested energy in \eqref{eq:energy_per_symbol_SU_2} is upper bounded as in \eqref{eq:convex_upperbound}. This proves $E_k(\tau, p_1, p_2) \leq \barE_{k} (\tau, p_1, p_2)$.

For a traditional radio communication system, when the receive SNR for CE at the ER is $\beta_k c_k$ which is the same as that for a backscatter communication system, it can be shown that the harvested energy is exactly $\barE_{k} (\tau, p_1, p_2)$ in \eqref{eq:convex_upperbound}, by following the steps in~\cite{YangHoGuan13}. This completes the proof.
\end{IEEEproof}
Numerical results in Section~\ref{Simulation} will show that this reduction in the harvested energy is marginal, which motivates the use of backscatter WET due to its low complexity at the ER.

\subsection{WET to Multiple ERs}\label{sec:WET_BF_MU}
To achieve WET to all ERs concurrently, we allow the use of multiple energy beams each toward one particular ER. Denote $\hatbA=[\hatba_1, \hatba_2, \cdots, \hatba_K]$. The beamformer is then chosen as a linear combination of the normalized estimated BS-CSI $\hatba_k$'s, i.e., 
\begin{align}
  \bw (\hatbA) = \sum_{k=1}^K \sqrt{\xi_k} \frac{\hatba_k}{\| \hatba_k \|_2}, \label{eq:beamformerMU}
\end{align}
where the weights $\xi_k$'s are subject to the condition $\sum_{k=1}^K \xi_k=1$.
\begin{Remark}\label{rem:asymptotic_optimality}
As the number of antennas $M$ tends to infinity, the beamformer in~\eqref{eq:beamformerMU} is asymptotically optimal~\cite{YangHoGuanMMIMO14}. This is the important motivation for choosing the beamformer in~\eqref{eq:beamformerMU}, while maintaining the flexibility of choosing different weights for energy beams toward different ERs.
\end{Remark}

Similar to the case of single ER, the energy harvested by ER $k$ is given by
\begin{align}\label{eq:energy_per_symbol_MU1}
   E_{k} = E_{k} (\tau, p_1, p_2, \xi_k) = p_2 (T-\tau) \bbE_{\bh_k, \hatbA} \left[\left|\bw^T (\hatbA) \bh_k \right|^2\right].
\end{align}
The harvested energy is given by the following lemma.
\begin{mylem}\label{lemma:HarvestedEnergy}
With the beamformer in~\eqref{eq:beamformerMU}, the harvested energy by ER $k$ is given by
\begin{align}\label{HarvestedEnergy0}
  E_{k} (\tau, p_1, p_2, \xi_k) &= p_2 M \beta_k \xi_k (T \!-\! \tau) \left( 1 \!-\! \frac{M \!-\! 1}{M} \bbE_{g_k}  \left[\frac{1}{\frac{\beta_k \tau p_1 |g_k|^2}{K M \sigma^2} \!+\! 1} \right] \right) \!+\! p_2 \beta_k (T \!-\! \tau) (1 \!-\! \xi_k).
\end{align}
\end{mylem}
\begin{IEEEproof}
See Appendix~\ref{AppendixEnergy_MU}.
\end{IEEEproof}
The first term in~\eqref{HarvestedEnergy0} is the harvested energy from the beam directly toward ER $k$, while the second term is the energy harvested from beams toward other ERs but still harvested by ER $k$.

From Lemma 2 in~\cite{YangHoGuanMMIMO14}, the harvested energy of a traditional radio communication system is the same as in \eqref{HarvestedEnergy0} for $g_k=1$. Hence, the backscatter communication system studied here generalizes the result of a traditional radio communication system.

We note that the harvested energy in \eqref{HarvestedEnergy0} appears analytically intractable. Therefore, we will obtain bounds for the harvested energy, to simplify analysis in subsequent sections. Before that, we give the following lemma.
\begin{mylem}\label{lem:Expectation}
  Let random variable $X \sim \calC \calN (0, \beta)$ and $c$ be some positive constant. Then,
  \begin{align}
    \bbE_X \left[ \frac{1}{1+ c |X|^2} \right] = \frac{\exp \left(\frac{1}{c \beta} \right) \Gamma_1 \left(\frac{1}{c \beta} \right) }{c \beta}, \label{eq:Expectation}
  \end{align}
\end{mylem}
\noindent where $\Gamma_1(t) \triangleq \int_{t}^{\infty} u^{-1} \exp(-u)  \text{d} u$ is an upper incomplete Gamma function.
Moreover, the expectation is lower and upper bounded as
\begin{align}
    \frac{\ln (1+2 c \beta)}{2 c \beta} < \bbE_X \left[ \frac{1}{1+ c |X|^2} \right] < \frac{\ln (1+c \beta)}{c \beta}. \label{eq:Expectation_Bound}
  \end{align}
\begin{IEEEproof}
The expectation in~\eqref{eq:Expectation} is obtained by standard integration. The lower bound and upper bound in~\eqref{eq:Expectation_Bound} is obtained from~\cite[(5.1.20)]{MathHandbook}.
\end{IEEEproof}

We assume the average power for each frame is $p_{\sf ave}$. Besides the energy consumption for channel training, all the remaining energy is used for WET. This implies the WET power is given as
\begin{align}
  p_2 = \frac{p_{\sf ave} T - \tau p_1}{T-\tau}. \label{eq:WET_power}
\end{align}
From~\eqref{eq:Expectation} and~\eqref{eq:WET_power}, the harvested energy in~\eqref{eq:energy_per_symbol_SU_2} is rewritten as
\begin{align}
  E_k (\tau, p_1, \xi_k) &=M \beta_k \xi_k (p_{\sf ave} T - \tau p_1) \left[ 1- \frac{M-1}{M} \frac{K M \sigma^2}{\beta_k^2 \tau p_1} \exp \left(\frac{K M \sigma^2}{\beta_k^2 \tau p_1} \right) \Gamma_1 \left(\frac{K M \sigma^2}{\beta_k^2 \tau p_1} \right) \right]+ \nonumber \\
   &\quad \;\; \beta_k (p_{\sf ave} T - \tau p_1) (1-\xi_k) \nonumber \\
   &\eqa \beta_k (p_{\sf ave} T - q) \left[ (M-1) \left[1- \frac{K M \sigma^2}{\beta_k^2 q} \exp \left(\frac{K M \sigma^2}{\beta_k^2 q} \right) \Gamma_1 \left(\frac{K M \sigma^2}{\beta_k^2 q} \right) \right] \xi_k +1 \right] \nonumber \\
   &\triangleq E_k(q, \xi_k), \label{lemma:HarvestedEnergy2}
\end{align}
where in (a) we perform algebraic simplification and introduce $q \triangleq \tau p_1$ which denotes the total energy used for channel training. We observe that the harvested energy depends only on $q$, since the CE time $\tau$ and the transmit power $p_1$ for CE are always coupled in~\eqref{lemma:HarvestedEnergy2}. Based on this observation, we will optimize the channel-training energy $q$ in next sections, instead of optimizing $\tau$ and $p_1$.

For the special case that $q=0$, no CSI is available and hence the ET performs omnidirectional transmission; then the harvested energy reduces to $E_k(0, \xi_k)=p_{\sf ave} T \beta_k$, as expected.

From Lemma~\ref{lem:Expectation} and~\eqref{lemma:HarvestedEnergy2}, we immediately obtain the bounds on the harvested energy as follows:
\begin{mylem}
  With the energy beamformer in~\eqref{eq:beamformerMU}, the energy harvested by ER $k$ is lower bounded by
  \begin{align}
      E_k (q, \xi_k)
   &> \beta_k (p_{\sf ave} T - q) \left[ (M-1) \left[1- \frac{K M \sigma^2}{\beta_k^2 q} \ln \left( 1+
   \frac{\beta_k^2 q}{K M \sigma^2} \right) \right] \xi_k +1 \right] \triangleq \tilE_k (q, \xi_k),   \label{eq:TotalEnergy_LB}
  \end{align}
  and it is upper bounded by
  \begin{align}
      E_k (q, \xi_k)
   &< \beta_k (p_{\sf ave} T - q) \left[ (M-1) \left[1- \frac{K M \sigma^2}{2\beta_k^2 q} \ln \left( 1+
   \frac{2\beta_k^2 q}{K M \sigma^2} \right) \right] \xi_k +1 \right]. \label{eq:TotalEnergy_UB}
  \end{align}
\end{mylem}
As will be numerically shown in Section \ref{Simulation}, the bounds in~\eqref{eq:TotalEnergy_LB} and~\eqref{eq:TotalEnergy_UB} are tight, especially for the lower bound. In the sequel, for analytical tractability, we take the lower bound $\tilE_k (q, \xi_k)$ as the energy harvested by ER $k$.

\section{Resource Allocation for WET in Backscatter Communication Systems} \label{sec:Formulation}
In this section, we further consider the resource allocation for WET in a backscatter communication system. We formulate a general optimization problem to maximize a total utility of the energy harvested by all ERs. Let $f_k(\tilE_k)$ be the utility of the $k$-th ER's harvested energy $\tilE_k$ given by \eqref{eq:TotalEnergy_LB}, assumed to be a monotonically increasing function of $\tilE_k$. Denote the vector of energy allocation weights for different energy beams by $\bm{\xi}=[\xi_1 \ \xi_2 \ \cdots \ \xi_K]^T$. The total utility is then given by \begin{align}
   U(q, \bm{\xi}) \triangleq \sum_{k=1}^K f_k \left(\tilE_k(q, \xi_k)\right). \label{eq:utility_func}
 \end{align}

We aim to maximize the total utility by optimizing the channel-training energy $q$ and the energy allocation weights $\bm \xi$ for different energy beams, subject to the total energy constraint (i.e., the training energy $q$ can not exceed the total energy $p_{\sf ave} T$ available in each frame) and the normalization constraint for the energy allocation weights $\bm{\xi}$. The general utility maximization problem is thus formulated as follows
\begin{subequations}\label{eq:optimP1}
\begin{align}
   \mathrm{(P_1)} \ \ \underset{ q, \ \bm{\xi} }{\text{max}} \ \
   & \sum_{k=1}^K f_k \left(\tilE_k(q, \xi_k)\right) \label{eq:rewardP1} \\
    \quad \text{s. t.} \ \
        &0 \leq q \leq p_{\sf ave} T \label{eq:const1P1} \\
        &\sum \limits_{k=1}^K \xi_k \leq 1\label{eq:const2P1} \\
        & \xi_k \geq 0, \; \forall k. \label{eq:const3P1}
\end{align}
\end{subequations}
In particular, we will investigate two utility maximization problems, namely, the weighted-sum-energy (WSE) maximization where the utility function $f_k(\tilE_k)=\theta_k  \tilE_k$ and the proportional-fair-energy (PFE) maximization where $f_k(\tilE_k)= \ln(\tilE_k)$.

\section{Optimal Solutions for Resource Allocation}\label{sec:solutions}
In this section, we obtain the optimal solutions for resource allocation for WET in a backscatter communication system, for the single-ER case and the multiple-ER case in Section \ref{Sec:Analysis_SU} and Section \ref{Sec:Analysis_MU}, respectively.
\subsection{Optimal Solution for WET to Single ER}\label{Sec:Analysis_SU}
In this section, we analyze the optimal solution for WET to single ER, i.e., $K=1, \xi_1=1$. From~\eqref{eq:TotalEnergy_LB}, the harvested energy by the ER is given by
\begin{align}
  \tilE_1 (q) &= M \beta_1 (p_{\sf ave} T - q) \left[ 1- \frac{\sigma^2 (M-1) \ln \left( 1+ \frac{\beta_1^2 q}{M \sigma^2} \right)}{\beta_1^2 q} \right]. \label{eq:TotalEnergy_SingleER}
\end{align}

We note that it suffices to maximize $\tilE_1(q)$, subject to $0 \leq q \leq p_{\sf ave} T$, since the utility function $f_1(\tilE_1(q))$ is assumed to be a monotonically increasing function of $\tilE_1(q)$, for both WSE maximization and PFE maximization. We obtain the optimal solution to Problem ($\mathrm{P_1}$) for WET to single ER, in the following theorem.

\begin{mythe}[Optimal Solution for WET to Single ER]\label{The:OptSolution_SingleUser}
The optimal solution to Problem ($\mathrm{P_1}$) for WET to single ER is given by
\begin{align}
  q^{\star}=\left\{ \begin{array}{cl}
  0, &\mbox{if}\; \sigma^2 \geq \frac{\beta_1 T p_{\sf ave} (M- 1)}{M} \\
  q_1^{\star}, \qquad &\mbox{otherwise} \\
  \end{array}
  \right. \label{eq:OptSolution_SU}
\end{align}
where $q_1^{\star}$ is the unique solution for $q \in (0, p_{\sf ave} T]$ to the equation
 \begin{align}
   \frac{p_{\sf ave}(M - 1)}{M} \left(\frac{\beta_1^2 q}{M \sigma^2}+1 \right) \ln \left( \frac{\beta_1^2 q}{M \sigma^2}+1 \right) - \frac{\beta_1^4 q^3}{M^2 \sigma^4 T} - \frac{\beta_1^2 q^2}{M T \sigma^2} - \frac{\beta_1^2 (M-1) p_{\sf ave} q}{M^2 \sigma^2} =0.\label{eq:OptSolution_SU_derivative}
 \end{align}
\end{mythe}

\begin{IEEEproof}
 (Sketch) When the noise variance $\sigma^2 \geq \frac{\beta_1 T p_{\sf ave} (M- 1)}{M}$, the objective function $\tilE_1 (q)$ can be easily shown to be monotonically decreasing with respect to $q$. The optimal solution is thus zero. When $\sigma^2 < \frac{\beta_1 T p_{\sf ave} (M- 1)}{M}$, the objective function $\tilE_1 (q)$ can be shown to be strictly concave with respect to $q$ for $q \in (0, p_{\sf ave} T]$. The solution is thus unique. See details in Appendix~\ref{app:SingleUserSolution}.
\end{IEEEproof}

Theorem~\ref{The:OptSolution_SingleUser} implies that when the noise level for CE at the ET is too high, it is better for the ET to broadcast energy in all directions, without beamforming. This is as expected, since the energy used for CE does not justify the gain achieved from beamforming.

\subsection{Optimal Solution for WET to Multiple ERs} \label{Sec:Analysis_MU}
In this section, we maximize the total utility for WET to multiple ERs. In particular, Section~\ref{Sec:Analysis_MU_WeightedSum} and Section~\ref{Sec:Analysis_MU_SumLog} consider the WSE maximization and the PFE maximization, respectively.

\subsubsection{Weighted-Sum-Energy Maximization} \label{Sec:Analysis_MU_WeightedSum}
For WSE maximization, the total utility is rewritten as
\begin{align}
U(q, \bm{\xi})
   &= (p_{\sf ave} T - q) \sum \limits_{k=1}^K \beta_k \theta_k + \sum_{k=1}^K \alpha_k (q) \xi_k,\label{eq:obj_P1_MU_WeightedSum}
\end{align}
where the function $\alpha_k(q)$ is given by
\begin{align}
  \alpha_k(q)= \theta_k \beta_k (M-1) (p_{\sf ave} T - q) \left[1- \frac{K M \sigma^2}{\beta_k^2 q} \ln \left( 1+
   \frac{\beta_k^2 q}{K M \sigma^2} \right) \right]. \label{eq:Constant_Alpha_q}
\end{align}
From the inequality $\ln (x+1) < x, \; \forall x > 0$, it is obvious that $\alpha_k(q) > 0$ for $q>0$. It is noted that $\alpha_k(q)$ is a strictly concave function of $q \in (0, p_{\sf ave} T)$.

We obtain the optimal solution to the WSE maximization problem in the following theorem.
\begin{mythe}\label{The:OptSolution_MU_WeightedSum}
The optimal energy allocation weights are
\begin{align}\label{eq:optimal_xi_WSE}
  \xi_k &= \left\{\! \! \!
  \begin{array}{cl}
    &1,  \quad k=k^{\star} \\
    &0, \quad \text{otherwise}
  \end{array}
  \right.
\end{align}
where $k^{\star}$ is given by the following criterion
\begin{align}
k^{\star} = \underset{k_1}{\arg \max} \quad \alpha_{k_1} (q_{k_1}^{\star}) + (p_{\sf ave} T - q_{k_1}^{\star}) \sum \limits_{k=1}^K \beta_k \theta_k, \nonumber
\end{align}
where $q_{k_1}^{\star}$ is the unique solution that maximizes the function $\alpha_{k_1}(q)$ for $q \in (0, p_{\sf ave} T]$. The corresponding optimal training energy $q^{\star} = q_{k^{\star}}^{\star}$.
\end{mythe}

\begin{IEEEproof}
From~\eqref{eq:obj_P1_MU_WeightedSum}, for any given $q \in \left(0, p_{\sf ave} T \right]$, the objective function is a linear function of $\xi_k$'s. To maximize the weighted sum of harvested energy, it suffices to allocate all energy to single energy beam toward the ER with the largest linear-combination weight $\alpha_k(q)$. This gives \eqref{eq:optimal_xi_WSE}. For $q \in (0, p_{\sf ave} T]$, the solution that maximizes the function $\alpha_{k_1}(q)$ is unique, as $\alpha_k(q)$ is strictly concave. This completes the proof.
\end{IEEEproof}

Theorem~\ref{The:OptSolution_MU_WeightedSum} implies that only one beam is used to transfer energy to one particular ER in the WET phase, although the ET has channel estimates for all ERs. Energy is thus wasted for estimating channels of other ERs. Moreover, when more antennas are employed at the ET, the only energy beam becomes more concentrated. Other ERs without dedicated energy beam can harvested very little energy, resulting in severe unfairness among ERs. This observation will be numerically verified in Section~\ref{Simulation}. Hence, we will consider another utility function that takes fairness among ERs into consideration, in the next section.

\subsubsection{Proportional-Fair-Energy Maximization} \label{Sec:Analysis_MU_SumLog}
In this section, we aim to maximize the log-sum of the energy harvested by all ERs, which is known to result in proportional fairness~\cite{FKelly97}.
From~\eqref{eq:TotalEnergy_LB}, the total utility is rewritten as
\begin{align}\label{eq:utility_PF}
  U(q, \bm{\xi}) &= \sum \limits_{k=1}^K \ln \left(b_k(q) \xi_k + d_k(q) \right),
\end{align}
where the quantities depending on $q$ are given by
\begin{align}
  b_k(q) &= \beta_k (M-1) (p_{\sf ave} T - q) \left[1- \frac{K M \sigma^2}{\beta_k^2 q} \ln \left( 1+
   \frac{\beta_k^2 q}{K M \sigma^2} \right) \right],
  \label{eq:PropFairCoeff_a_k}\\
  d_k(q) &= \beta_k \left( p_{\sf ave} T - q \right)\label{eq:PropFairCoeff_b_k}.
\end{align}

Hence, the PFE maximization problem is rewritten as
\begin{subequations}\label{eq:optimP2}
\begin{align}
   \mathrm{(P_2)} \ \ \underset{ q, \ \bm{\xi} }{\text{min}} \ \
   & - \sum \limits_{k=1}^K \ln \left(b_k(q) \xi_k + d_k(q) \right) \label{eq:rewardP2} \\
    \quad \text{s. t.} \ \ &\text{constraint} \; \eqref{eq:const1P1} , \eqref{eq:const2P1}, \eqref{eq:const3P1}
\end{align}
\end{subequations}
Before further analysis, we give the following definitions~\cite{GorskiBiconvex07}.
\begin{mydef}\label{def:biconvex_func}
  A function $g: \calX \times \calY \rightarrow \calR$ is called biconvex, if $g(x, y)$ is convex in $y$ for fixed $x \in \calX$ and is convex in $x$ for fixed $y \in \calY$.
\end{mydef}
\begin{mydef}\label{def:biconvex_problem}
  A problem is a biconvex problem, if it optimizes a biconvex function over a given biconvex or compact set.
\end{mydef}

We then have the following theorem for the Problem ($\mathrm{P_2}$).
\begin{mythe}\label{the:ProperFair_Biconvex}
  The Problem ($\mathrm{P_2}$) is a biconvex problem.
\end{mythe}

\begin{IEEEproof}
We note that the logarithm function is concave and increasing. Given $q$, the summation term in~\eqref{eq:rewardP2} is concave, as it is a composition of a concave and increasing function (i.e., $\ln(\cdot)$) and a concave function (i.e., linear function of $\xi_k$). The objective function~\eqref{eq:rewardP2} is thus strictly convex.

On the other hand, given $\bm{\xi}$, the argument of the logarithm function is concave, as $b_k(q)$ is concave (see Appendix~\ref{app:SingleUserSolution}). The objective function~\eqref{eq:rewardP2} is also strictly convex, as it is the sum of a family of compositions of a convex and decreasing function (i.e., $-\ln(\cdot)$) and a concave function (i.e., linear function of $\tilE_k(q)$). Clearly, the domain for Problem ($\mathrm{P_2}$) is a convex set. By Definition~\ref{def:biconvex_problem}, the Problem ($\mathrm{P_2}$) is a biconvex problem.
\end{IEEEproof}

In general, a biconvex problem is nonconvex and has multiple optima. Before giving the algorithm to find solution for the biconvex Problem ($\mathrm{P_2}$), we first decompose the problem into two subproblems. For Subproblem ($\mathrm{P_{2a}}$), given the training energy $q$, we optimize the energy allocation weights $\bm{\xi}$, namely, this performs beamforming energy allocation. For Subproblem ($\mathrm{P_{2b}}$), given the energy allocation weights $\bm{\xi}$, we optimize the training energy $q$, namely, this performs training energy allocation.

Subproblem ($\mathrm{P_{2a}}$) is a (strictly) convex optimization problem. There is thus a unique global solution. We give the optimal beamforming energy allocation for Subproblem ($\mathrm{P_{2a}}$), in Theorem~\ref{The:OptWeights_MU_PropFair}.

\begin{mythe}[Optimal Beamforming Energy Allocation]\label{The:OptWeights_MU_PropFair}
Given training energy $q \in (0, p_{\sf ave} T]$, the optimal solution for subproblem $\mathrm{(P_{2a})}$ follows a water-filling method. In particular, the optimal beamforming energy allocation weight is given by
\begin{align}
  \xi_k^{\star} (q) = \max \left\{0, \frac{1}{\nu^{\star}(q)} - \frac{d_k (q)}{b_k (q)} \right\}, \label{eq:OptXi}
\end{align}
where  the water-level $\frac{1}{\nu^{\star}(q)}$ is the unique solution to the equation
\begin{align}
\sum \limits_{k=1}^K \max \left\{0, \frac{1}{\nu(q)} - \frac{d_k (q)}{b_k (q)} \right\} = 1. \label{eq:Waterlevel}
\end{align}
\end{mythe}
\begin{IEEEproof}
It is proved by using Karush-Kuhn-Tucker (KKT) conditions. See Appendix~\ref{App:Water_Filling_Proof}.
\end{IEEEproof}

As shown in the proof for Theorem~\ref{the:ProperFair_Biconvex}, Subproblem $\mathrm{(P_{2b})}$ strictly convex. We can thus find the unique optimal solution $q_1^{\star} \in (0, p_{\sf ave} T]$, by using any convex optimization toolbox, such as \cite{CVXTool}, although it is difficult to obtain the closed-form solution.
We note that the objective function~\eqref{eq:rewardP2} is derived for WET via energy beamforming, and thus not applicable for the case of $q=0$ in which omnidirectional transmission is used. For that case, the total utility is a constant $\sum \nolimits_{k=1}^K \ln \left( \beta_k p_{\sf ave} T\right)$, regardless of the choose of $\xi_k$'s. For the case of $q = p_{\sf ave} T$, the total utility is $-\infty$, as no energy is harvested (i.e., no time is allocated for WET). Hence, the final optimal $q^{\star}$ for given $\bm{\xi}$ is chosen between $q_1^{\star}$ and zero.

For a biconvex problem, there is no algorithm that ensures to find the global optima\cite{GorskiBiconvex07}. In the state-of-the-art literature, the block coordinate descent (BCD) algorithm is computationally efficient and with performance guarantee, as it ensures to converge to a partial optimal solution~\cite{Tseng01}. Hence, we propose a BCD-based Algorithm \ref{Algorithm_PFE} as follows.
\begin{algorithm}[tb]
\caption{Algorithm for PFE Maximization:}
\label{Algorithm_PFE}
\begin{algorithmic}[1]
\vspace{.2cm}
\STATE Parameters: $\{\beta_k\}$, $p_{\sf ave}, T, M, K, \sigma^2, \epsilon$.
\STATE Initialization: Choose any $q_0 \in (0, p_{\sf ave} T]) $, and any feasible $\bm{\xi}_0$ that satisfies \eqref{eq:const2P1} and \eqref{eq:const3P1}, compute $U_0 = U(q_0, \bm{\xi}_0)$, set $U_1 = U_0 + 2 \epsilon, t=0$. \\
\WHILE{$|U_{t+1} - U_t| > \epsilon$}
		\STATE Keep $q_t$ fixed, use the water-filling results in Theorem~\ref{The:OptWeights_MU_PropFair} to find the optimal energy allocation weights $\bm{\xi}_{t+1}$.
		\STATE Keep $\bm{\xi}_{t+1}$ fixed, find $q_{t+1}$ that minimizes the objective function in~\eqref{eq:rewardP2}, by using standard convex optimization techniques.
		\STATE $t = t+1$.
		\STATE Use \eqref{eq:utility_PF} to compute the updated utility as $U_{t+1} = U(q_{t}, \bm{\xi}_{t})$.
\ENDWHILE

\RETURN $q^{\star} = q_{t}, \; \bm{\xi}^{\star} = \bm{\xi}_{t}, \; U_{\max}= U(q_t, \bm{\xi}_t)$.
\end{algorithmic}
\end{algorithm}


\section{Discussion}\label{Sec:Extension}
In this section, we discuss some extension work and interesting practical issues. First, the analysis and results in this paper are also \emph{applicable to WET in a time-division-multiplexing (TDM) manner}. The WET phase of $(T-\tau)$ symbol periods is divided into $K$ slots. The $k$-th WET slot consists of $\xi_k (T-\tau)$ successive symbol periods. The relative time allocation coefficients $\xi_k$'s are subject to $\sum \nolimits_{k=1}^K \xi_k = 1$. In the $k$-th WET slot, the ET delivers energy to ER $k$ via energy beamforming using only the estimated BS-CSI $\hatba_k$ for ER $k$. It can be shown that for WET in a TDM manner, the harvested energy is the same as in~\eqref{lemma:HarvestedEnergy2}, and all the resource allocation results are also the same as those for the current energy multicasting in this paper. Hence, the WET via energy multicasting is equivalent to WET in a TDM manner. This is basically because the amount of harvested energy has a linear relationship with either the energy-harvesting time or the power of the energy beam. In this paper, we use multiple beams to transfer energy to all ERs concurrently, as this operation can avoid frequent switches between WET to different ERs, in each frame.

Based on current work, other issues can be further addressed, such as WET under \emph{nonlinear} backscattering and \emph{channel reciprocity}. For ERs with {nonlinear} backscattering, the reflection coefficient $\rho_k$ varies with the incident power \cite{NonlinearBackscatter}, as a backscatter ER is typically matched at some fixed power level, and thus mismatched at other incident power levels. Given the ER hardware, the training energy can also be optimized, once the reflection characteristic is experimentally measured.

On the other hand, to achieve smaller size and lower cost, some RFID readers use single antenna for both transmission and reception, by introducing an RF isolator such as a circulator or a directional coupler\cite{NikitinRFID08}. {With the assumption of reciprocal channels, in \cite{Griffin11}, the phase of the forward channel of each reader antenna was estimated. However, there may be phase ambiguity of integer multiple of $\pi$, which may result into less power delivered to the RF tag. Nonlinear methods may achieve better estimation of forward channels, but are beyond the scope of this paper.}

\section{Numerical Results}\label{Simulation}
In this section, numerical simulations are given to corroborate our analysis. We assume that the average power is $p_{\sf ave} = 1$~W. We set the number of transmit antennas as $M=4$. We assume each frame consists of $T=200$ symbol periods. For convenience, we normalize each symbol period to be one second, resulting into 200 J of energy consumption in each frame. The carrier frequency is $5$~GHz, and the bandwidth is $100$~KHz. We set the power spectrum density of noise as $-140 \ \textrm{dBm/Hz}$, which implies the noise power at the ET is $\sigma^2 = -90$~dBm. We consider two ERs, i.e., $K=2$. We take the path loss model as $10^{-3} D^{-3}$, where the path loss exponent is 3, and $D$ is the distance between the ET and an ER. A $30$dB path loss is assumed at a reference distance of $1$~m. The energy harvesting efficiency at each ER is assumed to be $\eta = 0.8$. The reflection coefficient for both ERs is $\rho_1 = \rho_2= 0.8+0.5 i$. All the simulation is based on $100, 000$ Monte Carlo simulation runs.

\subsection{Single-ER Case}\label{sec:simulation_singleER}

We first consider the single-ER case. We assume the distance $D_1= 6$~m, which implies the path loss is $\beta_1 = 4.6296 \times 10^{-6}$. We first simulate the harvested energy for two benchmarks, i.e., the case of the perfect F-CSI and the case of no CSI. For both cases, no channel-training time nor energy is required. With the perfect F-CSI, the ET performs maximum-ratio-transmit (MRT), and the harvested energy is obtained as $2.96$~mJ. For the case of no CSI, the ET performs omnidirectional transmission without beamforming, and the harvested energy is $0.74$~mJ.

\begin{figure} [htbp]
\centering
\includegraphics[width=.7\columnwidth] {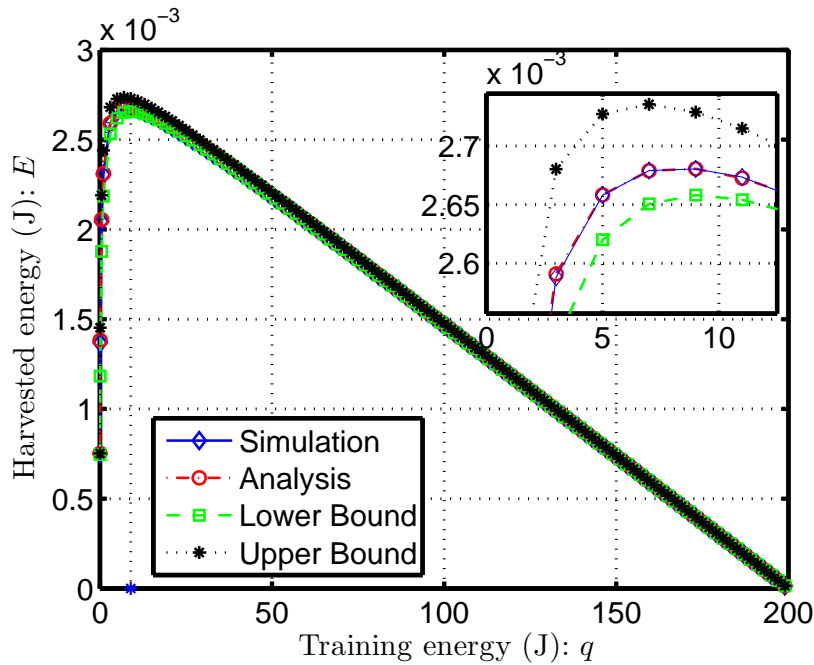}
\caption{Harvested energy v.s. training energy (single ER).}
\label{fig:Fig3}
\end{figure}

We then simulate the harvested energy by using the estimated BS-CSI. Fig.~\ref{fig:Fig3} plots the harvested energy versus the training energy $q$. {We observe that when the training energy approaches zero, the harvested energy is $0.75$ mJ, which approaches that for omnidirectional transmission. This is because no BS-CSI can be inferred, due to zero training energy.}

By simulation, the optimal training energy is $\hatq^{\star}=9.0$~J. The maximally harvested energy is $2.68$ mJ. From Theorem~\ref{The:OptSolution_SingleUser}, the optimal training energy is $q^{\star}=9.2$ J, which corroborates the simulation results. More importantly, we observe that compared to the MRT scheme, the maximally harvested energy via energy beamforming by using the estimated BS-CSI suffers only a \emph{slight} reduction of $9.6\%$. As expected, the harvested energy is also increased significantly by $262\%$, compared to omnidirectional transmission.
On the other hand, we observe that the dash-dot o-marker curve obtained by analysis coincides with the solid $\diamond$-marker curve obtained by simulations. We also see that the dashed $\square$-marker curve obtained by the lower bound is tight, which is obtained analytically.

\subsection{Comparison of Harvested Energy}\label{sec:simulation_comparison}
In this section, we compare the harvested energy to the net harvested energy by using the estimated F-CSI in a traditional radio communication system, to the harvested energy by using omnidirectional transmission, and to the harvested energy by using the perfect F-CSI. We assume that the traditional radio communication system operates in time-division-duplex mode. Thus, the ET has to first send forward pilots, such that the ER can estimate and feed back the F-CSI; additionally, the ET has to estimate backward channels by receiving pilots sent from the ER, and then recover the estimated F-CSI. For the ER, we consider the energy used for sending backward pilots and feeding back the estimated F-CSI, neglecting the energy used for data acquisition and computation. In particular, the ER employs analog feedback \cite{ChiuKimTWC08}, as it requires a lower feedback rate and has a smaller feedback delay. Following the scheme in Section II of \cite{ChiuKimTWC08}, the ET performs minimum-mean-square-error (MMSE) estimation for the backward channels, and uses the optimal MMSE filter to recover the estimated F-CSI. We denote the power for backward transmissions by $p_{\sf u}$. For the ER, the total energy used for backward transmissions is $E_0 = (M+1) p_{\sf u}$. In order to maximize the net harvested energy (after subtracting $E_0$), there is an optimal $p_{\sf u}$. In the sequel, the maximally net harvested energy is obtained by jointly optimizing both the forward resource allocation and $p_{\sf u}$.

\begin{figure}[htbp]
\centering
\includegraphics[width=.7\columnwidth] {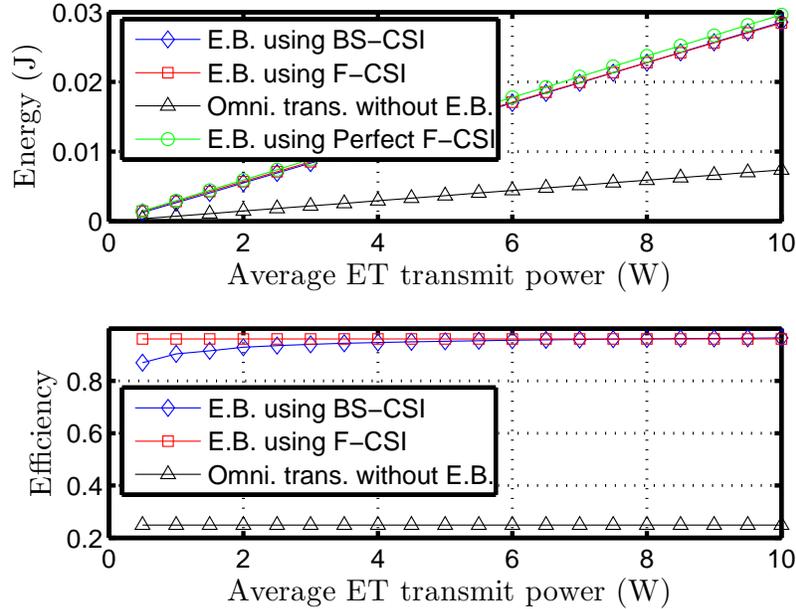}
 \caption{Harvested energy and efficiency for different energy beamforming (E.B.) schemes}
\label{fig:Fig6}
\end{figure}

{Fig. \ref{fig:Fig6} compares the harvested energy and the efficiency of different schemes in the upper and the lower figure, respectively. The efficiency is defined as the ratio of the harvested energy divided by the harvested energy by using the perfect F-CSI. As expected, the efficiency for omnidirectional transmission is $0.25$, as the harvested energy via energy beamforming by using the perfect F-CSI is increased by $M=4$ times, compared to that by using omnidirectional transmissions.}

We observe that when the average transmit power $p_{\sf ave}$ at the ET exceeds $4$ W, energy beamforming by using the estimated F-CSI and the estimated BS-CSI achieve the same efficiency\footnote{We observe that the efficiency by using the estimated F-CSI is almost constant. This is because for each $p_{\sf ave}$ at the ET, the estimated F-CSI at the ER is near perfect due to only one-way forward-channel propagation, and the optimal energy used for backward transmissions is numerically shown to be the same.} of about $96 \%$. For smaller $p_{\sf ave}$, the efficiency of energy beamforming by using the estimated BS-CSI still {exceeds $90\%$}, and is slightly lower than that by using the estimated F-CSI. For instance, for $p_{\sf ave} =2$ W, the harvested energy by using the estimated BS-CSI (i.e., $5.5$ mJ) suffers slight reduction, compared to that (i.e., $5.7$ mJ) by using the estimated F-CSI. After normalizing to the harvested energy (i.e., $5.92$~mJ) by using the perfect F-CSI, the efficiency of energy beamforming by using the estimated BS-CSI is $93\%$, which is slightly degraded, compared to the efficiency (i.e., $96\%$) by using the estimated F-CSI. This is encouraging, as almost all complexity of hardware and computation at the ERs is shifted to the ET, at the cost of slight reduction in the harvested energy. The energy beamforming by using the estimated BS-CSI is thus efficient and attractive for transferring energy to (ultra-)low-power and low-cost wireless devices.

\subsection{Multiple-ER Case}\label{sec:simulation_multipleER}
In this section, we simulate both the WSE and the PFE maximization problems, for the two-ER case. For WSE maximization, we fix $D_1 = 4$~m and $D_2 = 6$~m, and choose the combination weights $\theta_1=0.3$ and $\theta_2=0.7$, to balance the energy harvested by the nearer ER 1 and the further ER 2. From simulation, the optimal training energy is $q^{\star}=4.1$~J and the optimal weight is $\xi_1^{\star}=1$. The harvested energy is $9.71$ mJ and $0.7431$ mJ for ER 1 and for ER 2, respectively. From Theorem~\ref{The:OptSolution_MU_WeightedSum}, we have $q^{\star}=4.0, \; \xi_1^{\star}=1$. The harvested energy is $9.66$ mJ and $0.739$ mJ for ER 1 and for ER 2, respectively. The simulation results corroborate the analysis.

For PFE maximization, we fix $D_1=4$~m, and choose different distances $D_2 > D_1$. The results are given in~Table~\ref{table1}, in which we use the normal notation (e.g., $q$ and $\xi_k$) for the analytic results, and use the notations with hat ($\hatq$ and $\hat{\xi}_k$, respectively) for the numerical results. We observe that the obtained solution by using the BCD-based algorithm is close to the optimal solution. The simulations corroborate the analysis.
\begin{table} [htbp]
\centering
\caption{Optimal solutions versus distances (PFE Maximization)} \label{table1}
\small{
\begin{tabular}{*{11}{c}}
  \hline \hline
  $D_2$ & $q^{\star}$ & $q^{\star} \text{(BCD)}$ & $\hatq^{\star}$ & $\xi_1^{\star}$ & $\xi_1^{\star} \text{(BCD)}$ & $\hat{\xi}_1^{\star}$ & $E_1$ & $\hatE_1$ & $E_2$ & $\hatE_2$ \\
  \hline
  $4$   & 3.83 & 3.80  & 3.83 & 0.5000  & 0.4998 & 0.5001 & 6.052E-3 & 6.076E-3 & 5.908E-3 & 5.932E-3\\
  $5$   & 5.46 & 5.42  & 5.45 & 0.5112  & 0.5105  & 0.5102 & 6.040E-3 & 6.071E-3 & 2.915E-3 & 2.931E-3\\
  $6$   & 7.45 & 7.43  & 7.44 & 0.5285  & 0.5265  & 0.5248 & 6.148E-3 & 6.169E-3 & 1.563E-3 & 1.581E-3\\
  $7$  & 9.65  & 9.63  & 9.64 & 0.5502  & 0.5485 & 0.5476 & 6.238E-3 & 6.249E-3 & 9.006E-4 & 9.019E-4\\
  $8$  & 10.86  & 10.84  & 10.85 & 0.5898  & 0.5876 & 0.5862 & 6.490E-3 & 6.507E-3 & 5.299E-4 & 5.302E-4\\
  \hline
\end{tabular}
}
\end{table}

\begin{figure}[htbp]
\centering
\includegraphics[width=.7\columnwidth] {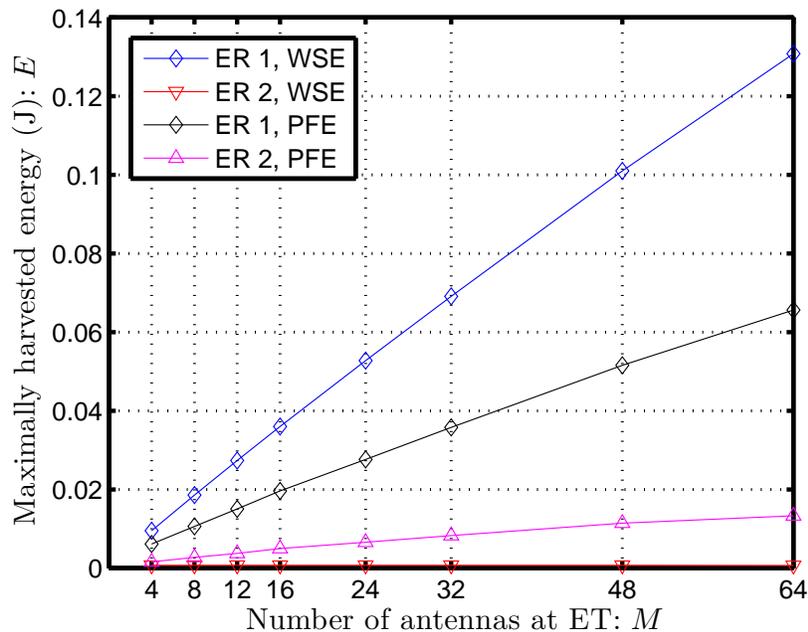}
\caption{Maximally harvested energy versus $M$.}
\label{fig:Fig5}
\end{figure}

Moreover, we compare the harvested energy for the WSE maximization problem and the PFE maximization problem, for different number of antennas $M$ deployed at the ET. We fix $D_1=4$~m, $D_2 = 6$~m. Fig. \ref{fig:Fig5} plots the maximally harvested energy of each ER versus $M$. For WSE maximization, we observe that as $M$ increases, the harvested energy by the nearer ER 1 increases, while the harvested energy by the further ER 2 remains as a small constant. This is because the ET uses only one energy beam toward the ER 1. Therefore, the harvested energy is unfair among ERs. For PFE maximization, however, the harvested energy by both ERs increases, as $M$ increases. Compared to WSE maximization, the harvested energy by the further ER 2 is increased significantly, although the energy harvested by the near ER 1 is about half of that for WSE maximization. The harvested energy is more balanced between the two ERs. Hence, we conclude that better fairness is achieved by PFE maximization.

\section{Conclusion} \label{Conclusion}
This paper studies the energy beamforming by using the estimated BS-CSI. We analyze the harvested energy, by investigating the effect of the ambiguity of backward channels. Moreover, we optimize the channel-training energy and the energy allocation weights, for two utility maximization problems. For WSE maximization, the optimal WET scheme is to use one energy beam, resulting in unfairness of harvested energy. For PFE maximization, we show that the problem is biconvex, and propose a BCD-based algorithm to find the close-to-optimal solution. {The harvested energy by using the estimated BS-CSI is numerically shown to suffer slight reduction, compared to that by using the perfect F-CSI, and also to the net harvested energy by using the estimated F-CSI in traditional radio communication systems.} Hence, the energy beamforming by using the estimated BS-CSI is a promising WET scheme, especially for transferring energy to (ultra-)low-power and low-cost wireless devices that are neither capable of estimating the channels nor sending pilots actively. Other interesting issues are also discussed and remain to be addressed, such as the WET under nonlinear backscattering and channel reciprocity.

\appendices
\section{Proof for Lemma~\ref{lemma:optimalbeamformerSU}}\label{appendix_proof_optimal_beamformer}
Recall the harvested energy given in~\eqref{eq:energy_per_symbol_2}.
The term in the round brackets of~\eqref{eq:energy_per_symbol_2} is the sum of a scaled identity matrix and
a rank-one matrix. The eigenvectors can be constructed as follows:
take the normalized $\frac{\hatbh_k}{\| \hatbh_k \|}$ as the right eigenvector corresponding to the maximal eigenvalue, and construct other mutually orthogonal eigenvectors by Gram-Schmidt algorithm. The term in the round brackets of~\eqref{eq:energy_per_symbol_2} is then maximized, when the beamformer is given by
  \begin{align}\label{eq:opt_beamformer_0}
    \bw_k (\hatba_k) =     \frac{\hatbh_k^{\ast}}{\| \hatbh_k \|}.
  \end{align}
Clearly, the argument of the inner expectation is still maximized, when a common angle $\angle g_k$ is introduced to all beamforming weights. Hence, an optimal beamformer that depends on the available estimate of the backscatter channel is given by
  \begin{align}\label{eq:opt_beamformer_1}
    \bw_k (\hatba_k) = \frac{\hatbh_k^{\ast} e^{- j \angle g_k}}{\| \hatbh_k \|} =     \frac{\hatba_k^{\ast} }{\| \hatba_k \|}.
  \end{align}

With the beamformer in~\eqref{eq:opt_beamformer_1},
the maximally harvested energy in~\eqref{eq:energy_per_symbol_2}
is rewritten by
\begin{align}\label{eq:energy_per_symbol_SU2}
    E_{0,k} (\tau, p_1) &= p_2 \bbE_{g_k} \Big[
    \frac{\beta_k \sigma_{e, k}^2 (g_k)}{\beta_k + \sigma_{e, k}^2 (g_k)}
    + \frac{\beta_k^2 \bbE_{\hatbh_k | g_k} \left[ \hatbh_k^H \hatbh_k \right]}{\left(\beta_k + \sigma_{e, k}^2 (g_k) \right)^2 } \Big] \eqa p_2 \beta_k \bbE_{g_k} \bigg[
    \frac{M \beta_k + \sigma_{e, k}^2 (g_k)}{\beta_k + \sigma_{e, k }^2 (g_k)} \bigg] \nonumber \\
    &\eqb p_2 M \beta_k \left( 1- \frac{M-1}{M} \bbE_{g_k}  \left[\frac{1}{\frac{\beta_k \tau p_1 |g_k|^2}{K M \sigma^2} + 1} \right] \right)
\end{align}
where (a) is obtained by substituting $\bbE_{\hatbh_k | g_k} \left[ \hatbh_k^H \hatbh_k \right] = M \left(\beta_k + \sigma_{e, k}^2 (g_k)\right)$, which is from the fact that the estimated F-CSI is conditionally distributed as $\hatbh_k | g_k \sim \calC \calN \left( \mathbf{0}_M, \left( \beta_k + \sigma_{e, k}^2 (g_k) \right) \bI_M  \right)$, (b) is from the conditional error variance given by~\eqref{eq:CondErrorVar}.

\section{Proof of Lemma~\ref{lemma:HarvestedEnergy}}\label{AppendixEnergy_MU}
Substituting~\eqref{eq:beamformerMU} into~\eqref{eq:energy_per_symbol_MU1}, the harvested energy by ER $k$ is rewritten as
\begin{align}
  E_{k} &(\tau, p_1, p_2, \xi_k) = \nonumber \\
  &p_2 (T-\tau) \Bigg( \bbE_{\hatbA} \bbE_{\bh_k \left| \hatbA \right.} \bigg[ \frac{\xi_k \left| \bh_k^H \hatba_k \right|^2}{\| \hatba_k \|^2} \bigg] +
  \bbE_{\hatbA} \bbE_{\bh_k \left| \hatbA \right.} \bigg[ \frac{ \sqrt{\xi_k} \left(\bh_k^H \hatba_k \right)^H }{\| \hatba_k \|_2}  \sum_{i \neq k} \frac{\sqrt{\xi_i} \bh_k^H \hatba_i}{\| \hatba_i \|_2} \bigg] \label{EHEnergyA} \\
  &+ \bbE_{\hatbA} \bbE_{\bh_k \left| \hatbA \right.} \bigg[ \frac{ \sqrt{\xi_k} \bh_k^H \hatba_k} {\| \hatba_k \|_2}  \sum_{i \neq k} \frac{\sqrt{\xi_i} \left(\bh_k^H \hatba_i \right)^H}{\| \hatba_i \|_2} \bigg] +
  \bbE_{\hatbA} \bbE_{\bh_k \left| \hatbA \right.}  \bigg[ \sum_{j \neq k} \sum_{l \neq k} \frac{ \sqrt{\xi_j \xi_l} \left(\bh_k^H \hatba_j \right)^H \bh_k^H \hatba_l}{\| \hatba_j \|_2 \| \hatba_l \|_2} \bigg] \Bigg). \nonumber
\end{align}

In the sequel, we investigate the four terms in the outer round bracket of~\eqref{EHEnergyA}. Given $g_k$, the random variable $\hatbh_k \triangleq \frac{\hatba_k}{g_k} = \bh_k + \be_k$. Recall that conditioned on $\hatba_k$ and $g_k$, the distribution of $\bh_k$ is given by~\eqref{eq:conditional_distribution}. Conditioned on the channel estimate $\hatbA$, the first term is rewritten as
\begin{align}
  \bbE_{\hatbA} \bbE_{\bh_k \left| \hatbA \right.} \bigg[ \frac{\xi_k \left| \bh_k^H \hatba_k \right|^2}{\| \hatba_k \|^2} \bigg] &= \xi_k \bbE_{g_k} \bbE_{\hatba_k | g_k} \bigg[ \frac{\hatba_k^H \bbE_{\bh_k \left| \hatba_k, g_k \right.} \left[\bh_k \bh_k^H \right] \hatba_k}{\| \hatba_k \|^2} \bigg]  \nonumber \\
  &\eqa M \beta_k \xi_k \left( 1- \frac{M-1}{M} \bbE_{g_k}  \left[\frac{1}{\frac{\beta_k \tau p_1 |g_k|^2}{K M \sigma^2} + 1} \right] \right) .\label{EHEnergyA1}
\end{align}
where (a) is obtained by following similar steps in the proof (Appendix~\ref{appendix_proof_optimal_beamformer}) for Lemma~\ref{lemma:optimalbeamformerSU}.

The second term in~\eqref{EHEnergyA} is rewritten as
\begin{align}
  \bbE_{\hatbA} &\bbE_{\bh_k \left| \hatbA \right.} \bigg[ \frac{ \sqrt{\xi_k} \left(\bh_k^H \hatba_k \right)^H }{\| \hatba_k \|_2}  \sum_{i \neq k} \frac{ \sqrt{\xi_i} \bh_k^H \hatba_i}{\| \hatba_i \|_2} \bigg]
  = \sqrt{\xi_k} \sum_{i \neq k}  \sqrt{\xi_i} \bbE_{g_k} \bbE_{\hatba_k, \hatba_i | g_k} \left[ \frac{\hatba_k^H \left(\bbE_{\bh_k | \hatba_k, g_k} \left[ \bh_k \bh_k^H \right] \right)\hatba_i}{\left\|\hatba_k\right\| \left\|\hatba_i\right\|}\right]  \nonumber \\
&\eqa \sqrt{\xi_k} \sum_{i \neq k}  \sqrt{\xi_i} \bbE_{g_k} \bbE_{\hatba_k, \hatba_i | g_k} \left[ \frac{\hatba_k^H}{\left\|\hatba_k\right\|}  \left( \frac{\beta_k \sigma_{e, k}^2 (g_k)}{\beta_k + \sigma_{e, k}^2 (g_k)} \bI_M + \frac{\beta_k^2 \hatba_k \hatba_k^H}{|g_k|^2 \left(\beta_k + \sigma_{e, k}^2 (g_k) \right)^2 }\right) \frac{\hatba_i}{\left\|\hatba_i\right\|}\right] \nonumber \\
   &\eqb 0,  \label{EHEnergyA2}
\end{align}
where ($a$) is from~\eqref{eq:conditional_distribution}, and (b) is duo to the fact that $\frac{\hatba_k}{\left\|\hatba_k\right\|}$ and $\frac{\hatba_i}{\left\|\hatba_i\right\|}$ are independent zero-mean random vectors, for any $i \neq k$.

The third term in~\eqref{EHEnergyA}, which is the conjugate of the second term in~\eqref{EHEnergyA}, is similarly obtained as
\begin{align}
  \bbE_{\hatbA} \bbE_{\bh_k \left| \hatbA \right.} \bigg[ \frac{ \sqrt{\xi_k} \bh_k^H \hatba_k} {\| \hatba_k \|_2}  \sum_{i \neq k} \frac{\sqrt{\xi_i} \left(\bh_k^H \hatba_i \right)^H}{\| \hatba_i \|_2} \bigg] &=0. \label{EHEnergyA3}
\end{align}

The fourth term in~\eqref{EHEnergyA} is rewritten as
\begin{align}
  \bbE_{\hatbA} &\bbE_{\bh_k \left| \hatbA \right.}  \bigg[ \sum_{j \neq k} \sum_{l \neq k} \frac{ \sqrt{\xi_j \xi_l} \left(\bh_k^H \hatba_j \right)^H \bh_k^H \hatba_l}{\| \hatba_j \|_2 \| \hatba_l \|_2} \bigg] = \sum_{j \neq k} \sum_{l \neq k} \sqrt{\xi_j \xi_l} \bbE_{\hatbA} \left[\frac{\hatba_j^H \left( \bbE_{\bh_k | \hatba_k} \left[\bh_k \bh_k^H \right] \right) \hatba_l}{\left\|\hatba_j \right\| \left\|\hatba_l \right\|}\right] \nonumber \\ &\eqa \sum_{j \neq k} \xi_j \bbE_{\hatba_k, \hatba_j} \left[\frac{\hatba_j^H \left( \bbE_{\bh_k | \hatba_k} \left[\bh_k \bh_k^H \right] \right) \hatba_j}{\left\|\hatba_j \right\|^2}\right] \nonumber \\
  &\eqb \sum_{j \neq k} \xi_j \bbE_{\hatba_k, \hatba_j} \left[\frac{\hatba_j^H \left( \bbE_{g_k} \left[ \frac{\beta_k \sigma_{e, k}^2 (g_k)}{\beta_k + \sigma_{e, k}^2 (g_k)} \bI_M + \frac{\beta_k^2 \hatba_k \hatba_k^H}{|g_k|^2 \left(\beta_k + \sigma_{e, k}^2 (g_k) \right)^2 }\right] \right) \hatba_j}{\left\|\hatba_j \right\|^2}\right] \nonumber \\
  &\eqc \sum_{j \neq k} \xi_j \beta_k
  \eqd \beta_k (1-\xi_k)
    \label{EHEnergyA4}
\end{align}
where (a) is from the fact that $\frac{\hatba_j}{\left\|\hatba_j\right\|}$ and $\frac{\hatba_l}{\left\|\hatba_l\right\|}$ are independent zero-mean random vectors, for any $j \neq l$, (b) is from~\eqref{eq:conditional_distribution}, (c) is from the fact that conditioned on $g_k$, the vector $\hatba_k$ is distributed as $\calC \calN \left(\mathbf{0}_M, |g_k|^2 \left(\beta_k + \sigma_{e, k}^2 (g_k)\right) \right)$, and (d) comes from the normalization condition $\sum \nolimits_{j=1}^K \xi_j =1$.

Substituting~\eqref{EHEnergyA1},~\eqref{EHEnergyA2},~\eqref{EHEnergyA3} and~\eqref{EHEnergyA4} into~\eqref{EHEnergyA}, we obtain the harvested energy as in~\eqref{HarvestedEnergy0}.

\section{Proof of Theorem~\ref{The:OptSolution_SingleUser}} \label{app:SingleUserSolution}
For convenience, the objective function in~\eqref{eq:TotalEnergy_SingleER} is rewritten as
\begin{align}
  \tilE_1(x) = M \beta_1 T (p_{\sf ave} - c_1 x) \left[ 1-  \frac{c_2 \ln ( x+1)}{x} \right],
\end{align}
where $c_1 = \frac{M \sigma^2}{\beta_1^2 T}$, the introduced variable $x = \frac{ \beta_1^2 q}{M \sigma^2} \in \left(0, {p_{\sf ave}}{c_1^{-1}} \right]$, and $c_2 = \frac{M - 1}{M} \in (0, 1)$.

The first derivative is derived as follows
\begin{align}\label{eq:first_derivative}
  \tilE_1^{\prime} (x) =  \frac{M \beta_1 T}{x^2 (x+1)} \left[ c_2 p_{\sf ave} (x+1) \ln(x+1) - c_1 x^3 -c_1 (1-c_2) x^2 - c_2 p_{\sf ave} x \right].
\end{align}

From~\eqref{eq:first_derivative}, the second derivative is then derived as follows
\begin{align}\label{eq:second_derivative}
  \tilE_1^{\prime \prime} (x) &=  - \frac{c_2 p_{\sf ave} M \beta_1 T}{x^2 (x+1)} \left[ \frac{2 (x+1) \ln(x+1)}{x} - \frac{x}{x+1} \left(1-\frac{c_1 x}{p_{\sf ave}}\right)- 2 \right] \nonumber \\
  &\eqa - \frac{c_2 p_{\sf ave} M \beta_1 T}{x^2 (x+1)} \left[ \frac{2 (x+1)}{x} \sum \limits_{n=1}^{\infty} \frac{1}{n} \left(\frac{x}{x+1}\right)^n - \frac{x}{x+1} \left(1-\frac{c_1 x}{p_{\sf ave}}\right)- 2 \right] \nonumber \\
  &= - c_2 p_{\sf ave} M \beta_1 T \left[ \sum \limits_{n=3}^{\infty} \frac{x^{n-3}}{n (x+1)^n} + \frac{c_1}{p_{\sf ave} (x+1)^2}\right] <0
\end{align}
where (a) is from the Taylor's expansion $\ln (x+1) = \sum \nolimits_{n=1}^{\infty} \frac{1}{n} \left(\frac{x}{x+1}\right)^n, \; \forall x > - \frac{1}{2}$.

From L$^{\prime}$H\^{o}pital's rule, the right limit at $x=0$ for the first derivative function is given by 
\begin{align}
  \lim \limits_{x \rightarrow 0^{+}} \tilE_1^{\prime} (x) & \!=\! M \beta_1 T \lim \limits_{x \rightarrow 0^{+}} \frac{c_2 p_{\sf ave} \!-\! (6 c_1 x-2c_1 + 2 c_1 c_2)(x+1)}{6x+2} \!=\! \frac{M \beta_1 T \left[c_2 p_{\sf ave} \!-\! 2 c_1(1-c_2) \right]}{2}  \label{eq:right_limit_at_zero}
\end{align}
We consider two scenarios. First, when $\sigma^2 < \frac{\beta_1 T p_{\sf ave} (M- 1)}{M} $ (i.e., $c_1 < \frac{c_2 p_{\sf ave}}{2 (1- c_2)}$), from~\eqref{eq:right_limit_at_zero}, we have $\lim \limits_{x \rightarrow 0^{+}} \tilE_1^{\prime} (x) > 0$. There thus exists a unique solution $x^{\star} \in \left(0, {p_{\sf ave}}{c_1^{-1}}\right]$, since $\tilE_1 (x)$ is concave with respect to $x$ and $\tilE_1^{\prime} ({p_{\sf ave}}{c_1^{-1}}) =0$.


Second, when $\sigma^2 \geq \frac{\beta_1 T p_{\sf ave} (M- 1)}{M} $ (i.e., $c_1 \geq \frac{c_2 p_{\sf ave}}{2 (1- c_2)}$), from~\eqref{eq:right_limit_at_zero}, we have $\lim \limits_{x \rightarrow 0^{+}} \tilE_1^{\prime} (x) \leq 0$. Moreover, from the Taylor's expansion of $\ln (x+1) $, the first derivative in~\eqref{eq:first_derivative} is rewritten as
\begin{align}\label{eq:first_derivative_2}
  \tilE_1^{\prime} (x)
  &=  \frac{M \beta_1 T}{x+1} \left[ \frac{c_2 p_{\sf ave}}{2 (x+1)}+c_2 p_{\sf ave} \sum \limits_{n=3}^{\infty} \frac{1}{n} \frac{x^{n-2}}{(x+1)^{n-1}} - c_1 x -c_1 (1-c_2) \right] \nonumber \\
  &\lea  \frac{M \beta_1 T x}{x+1} \left[ c_2 p_{\sf ave} \sum \limits_{n=3}^{\infty} \frac{1}{n} \frac{x^{n-3}}{(x+1)^{n-1}} - c_1\right] \nonumber \\
  &\leb  \frac{M \beta_1 T c_1}{ x (x+1)} g(x)
\end{align}
where (a) is from the fact $c_2 p_{\sf ave} \leq c_1 (1 - c_2)$ and $x \geq 0$,  (b) is due to the fact $c_2 p_{\sf ave} \leq c_1 (1 - c_2)$, and the function $g(x)$ is given by \begin{align}
  g(x) \triangleq \left[ 2 (1 - c_2) \sum \limits_{n=3}^{\infty} \frac{1}{n} \frac{x^{n-1}}{(x+1)^{n-1}} - x^2\right].
\end{align}
Clearly, we have $g(0)=0$. The derivative of $g(x)$ is derived as
\begin{align}
  g^{\prime}(x) &= \frac{2 (1 - c_2) x}{x+1} \sum \limits_{n=3}^{\infty} \frac{(n-1) x^{n-1}}{n (x+1)^{n-1}} - 2 x \nonumber \\
  &\lea \frac{4 (1 - c_2)}{3} \sum \limits_{n=3}^{\infty} \frac{x^{n-1}}{(x+1)^{n-1}} - 2 x = x \left[ \frac{4 (1 - c_2)}{3 (x+1)^2}-2 \right]
  \leb 0
\end{align}
where (a) is from $x \geq 0$, and (b) is from $0 < c_2 < 1$ and $x \geq 0$. The derivative in~\eqref{eq:first_derivative_2} is always non-positive, due to $g(x) \leq 0,\; \forall x \in \left(0, {p_{\sf ave}}{c_1^{-1}}\right]$. Hence, when $\sigma^2 \geq \frac{\beta_1 T p_{\sf ave} (M- 1)}{M} $, the optimal $x^{\star} = 0$. For $\sigma^2 < \frac{\beta_1 T p_{\sf ave} (M- 1)}{M}$, the optimal $x^{\star}$ is the unique solution for $\tilE_1^{\prime} (x) =0$, where the first derivative is given by~\eqref{eq:first_derivative}.
This completes the proof.

\section{Proof for Theorem~\ref{The:OptWeights_MU_PropFair}} \label{App:Water_Filling_Proof}
For notational convenience, we omit the given argument $q$ in this proof. The Lagrangian is constructed as follows
\begin{align}
  L(\xi_1, \cdots, \xi_K, \lambda_1, \cdots, \lambda_K, \nu)= -\sum \limits_{k=1}^K \ln (b_k \xi_k + d_k) - \sum \limits_{k=1}^K \lambda_k \xi_k + \nu \left( \sum \limits_{k=1}^K\xi_k - 1 \right).
\end{align}

The Karush-Kuhn-Tucker (KKT) conditions are thus written as
\begin{align}
  &\xi_k^{\star} \geq 0, \; \forall k \label{eq:KKT_con1}\\
  &\sum \limits_{k=1}^K\xi_k^{\star} = 1 \label{eq:KKT_con2}\\
  &\lambda_k^{\star} \geq 0, \; \forall k \label{eq:KKT_con3}\\
  &\lambda_k^{\star} \xi_k^{\star} = 0 \label{eq:KKT_con4}\\
  &-\frac{b_k}{b_k \xi_k^{\star} + d_k} - \lambda_k^{\star} + \nu^{\star} \eqa 0, \; \forall k \label{eq:KKT_con5}
\end{align}
where the equality (a) is derived by taking the first derivative of the Lagrangian. We note that $\lambda_k^{\star}$ is a slack variable which can be eliminated. The conditions~\eqref{eq:KKT_con3}~\eqref{eq:KKT_con4}~\eqref{eq:KKT_con5}
are thus rewritten as
\begin{align}
  &\xi_k^{\star} \left( \nu^{\star} - \frac{b_k}{b_k \xi_k^{\star} + d_k}\right)= 0 \label{eq:KKT_con4b}\\
  &\nu^{\star} \geq \frac{b_k}{b_k \xi_k^{\star} + d_k}, \quad \forall k \label{eq:KKT_con3b}
\end{align}

If $\nu^{\star} < \frac{b_k}{d_k}$, the condition~\eqref{eq:KKT_con3b} can only hold for $\xi_k^{\star} > 0$. Then the condition~\eqref{eq:KKT_con4b} implies $\nu^{\star} = \frac{b_k}{b_k \xi_k^{\star} + d_k}$. Equivalently, we have $\xi_k^{\star} = \frac{1}{\nu^{\star}} - \frac{d_k}{b_k}$. If $\nu^{\star} \geq \frac{b_k}{d_k}$, then $\xi_k^{\star} > 0$ is impossible. This is because the complementary slackness condition~\eqref{eq:KKT_con4b} is violated, as $\nu \geq \frac{b_k}{b_k \xi_k^{\star} + d_k}$, if $\xi_k^{\star} > 0$. That is, we have $\xi_k^{\star}=0$, when $\nu^{\star} \geq \frac{b_k}{d_k}$. Hence, we obtain the optimal WET time allocation coefficients
\begin{align}
  \xi_k^{\star} = \max \left\{0, \frac{1}{\nu^{\star}} - \frac{d_k}{b_k} \right\}. \label{eq:OptXi}
\end{align}
Substituting~\eqref{eq:OptXi} into the condition~\eqref{eq:KKT_con2}, we obtain
\begin{align}
\sum \limits_{k=1}^K \max \left\{0, \frac{1}{\nu^{\star}} - \frac{d_k}{b_k} \right\} = 1. \label{eq:Waterlevel}
\end{align}
The left hand side of~\eqref{eq:Waterlevel} is a piece-wise linear increasing function of $\frac{1}{\nu^{\star}}$, with breaking point at $\frac{d_k}{b_k}$. The equation~\eqref{eq:Waterlevel} thus has a unique solution that is readily determined. This completes the proof.

\renewcommand{\baselinestretch}{1.45}
\bibliography{IEEEabrv,reference1503}
\bibliographystyle{IEEEtran}

\end{document}